\documentclass[12pt]{iopart}

\usepackage{graphicx}
\usepackage[usenames]{color}
\newcommand{\beq}{\begin{equation}}
\newcommand{\eeq}{\end{equation}}
\newcommand{\beqa}{\begin{eqnarray}}
\newcommand{\eeqa}{\end{eqnarray}}
\newcommand{\ba}{\begin{array}}
\newcommand{\ea}{\end{array}}

\begin{document}

\title{Self-trapping of a binary Bose-Einstein condensate 
induced by interspecies interaction}
\author{S. K. Adhikari
} 
\address{Instituto de
F\'{\i}sica Te\'orica, UNESP - Universidade Estadual Paulista,
01.140-070 S\~ao Paulo, S\~ao Paulo, Brazil
}

\begin{abstract}

The problem of self-trapping of a Bose-Einstein condensate (BEC) and a 
binary BEC in an optical lattice (OL) and double well (DW)  is 
studied using the mean-field Gross-Pitaevskii equation. For both DW and OL, 
permanent self-trapping occurs in a window  
of the repulsive nonlinearity $g$ of the GP equation: $g_{c1}<g<g_{c2}$. 
In case of OL, the critical nonlinearities $g_{c1}$ and $g_{c2}$ 
correspond to a window of chemical potentials $\mu_{c1}<\mu<\mu_{c2}$ 
defining the band gap(s) of the periodic OL. The permanent self-trapped 
BEC in an OL usually represents a breathing oscillation of a stable 
stationary gap soliton. The permanent self-trapped BEC in a DW, 
on the other hand, is a dynamically stabilized state without any 
stationary counterpart. For a binary BEC with intraspecies 
nonlinearities outside this window of nonlinearity, a permanent self 
trapping can be induced by tuning the interspecies interaction such that 
the effective nonlinearities of the components fall in the above window. 
\end{abstract}

\pacs{03.75.Lm, 03.75Kk, 03.75.Mn}

\maketitle

\section{Introduction}

\label{I}

{After the experimental observation of a Bose-Einstein condensate (BEC), 
it is realized that a quasi one-dimensional (1D) cigar-shaped trap 
\cite{gor} is convenient to study many novel and complex phenomena. In a 
cigar-shaped BEC, apart from a harmonic trap, double well (DW) 
\cite{GatiJPB2007}, periodic optical-lattice (OL) \cite{kast}, and 
quasi-periodic bichromatic OL \cite{bich} traps have been used.  {
Usually, a 
repulsive BEC is localized in laboratory in an infinite trap. However, 
several other types of counter-intuitive localization of a repulsive BEC 
in a trap of finite height, where possible Josephson tunneling 
\cite{GatiJPB2007,cataliotti,java} of quantum fluids through barriers of 
finite height is expected to lead to delocalization, have lately drawn 
much attention. Of these, self trapping or localization of a repulsive 
BEC predominantly in one site of a DW \cite{trappingPRL,DW,DW1,hanpu,milburn}, 
triple well \cite{liu} and OL \cite{GatiJPB2007,ol1,ol2,strap} potential has 
been the subject matter of many investigations. Macroscopic self 
trapping of a BEC was first predicted theoretically 
\cite{java,trappingPRL,OstrovskayaPRA2000} and then observed 
experimentally \cite{GatiJPB2007,cataliotti}. It is generally believed 
that self trapping is an intrinsic {\it dynamical} phenomenon, without 
any stationary counterpart. In this connection we note that the Anderson 
localization \cite{bich,ander} in a quasi-periodic or random potential 
and a gap \cite{gap} soliton in a periodic potential are both  
{\it stationary} states of the system. In the present critical study of 
self trapping in DW and OL potentials we find that, although, in the 
former case, is is a {\it dynamical} phenomenon, in the latter case, 
contrary to general belief, localization takes place in a {\it 
stationary} gap soliton state. There is no symmetry broken stationary 
state corresponding to the self-trapped state in DW.


To understand the self trapping in an OL and a DW potential, we perform 
extensive numerical simulation of self trapping of a BEC using the 
solution of the Gross-Pitaevskii (GP) equation. In our study, we find a 
striking similarity between self trappings in OL and DW. Both occur in a 
window of repulsive nonlinearity $g$ of the GP equation: 
$g_{c1}<g<g_{c2}$. The numerical values of $g_{c1}$ and $g_{c2}$ depend 
on the respective trap parameters. (The existence of the upper limit of 
nonlinearity $g_{c2}$ was never noted in previous studies.)

Although, the self trapping of a repulsive BEC in a DW represents a 
relatively simple mathematical problem, well described by the analytic 
two-mode model \cite{trappingPRL,milburn}, the self trapping of a 
repulsive BEC in a periodic OL involving an infinite number of wells, on 
the other hand, poses a formidable mathematical problem and the 
fundamental mechanism in this case is not well understood. For a 
repulsive BEC in an OL, one has localized gap-soliton states 
\cite{kast,gap,gap2,gap3}. We demonstrate that, in an OL, self trapping 
represents a permanent breathing oscillation of a stable stationary gap 
soliton. For self trapping in OL, we consider compact state(s) localized 
mostly on a single site of OL, in contrast to previous considerations of 
self trapping on multiple OL sites \cite{ol1,ol2,gapself}. In the case 
of an OL, the critical nonlinearities $g_{c1}$ and $g_{c2}$ for self 
trapping define the window of chemical potentials corresponding to the 
band gap(s).
}

We also consider the self trapping of a binary BEC in OL and DW with 
tunable interspecies interaction near a Feshbach resonance \cite{fesh}.  
For zero interspecies interaction, self trapping takes place if the 
nonlinearities $g_i, i=1,2$ of the components are in the window 
$g_{c1}<g_i<g_{c2}$, and there is no self trapping if $g_i$'s are 
outside this window.  In the presence of interspecies interaction, the 
effective nonlinearity of both components (arising from a combination of 
inter and intraspecies interactions) should lie in the above window for 
self trapping. When both $g_1,g_2 > g_{c2}$, for zero interspecies 
interaction there is no self trapping. We illustrate, using the solution 
of the GP equation, that, for both OL and DW, if $g_{12}$ has an 
adequate attractive (negative) value, the effective nonlinearity of each 
component is reduced and one could have permanent self trapping. When 
both $g_1,g_2 < g_{c2}$, for zero interspecies interaction there is no 
self trapping. In this case, an appropriate repulsive (positive) 
interspecies nonlinearity can lead to effective nonlinearities of the 
components in the above window, and hence result in permanent self 
trapping.

\section{Analytical formulation for Double-Well (DW) potential}

\label{IIB}

We consider a binary cigar-shaped BEC 
in a DW. Atoms of both species (in two different hyperfine states) 
are assumed to have mass  $m$ and number $N$
and the DW acts 
in the axial $\hat x$ direction. 
Starting with the system of coupled 3D GP 
equations of the binary BEC one can reduce them to the following 
(dimensionless) 1D equations \cite{cigar}:
\begin{eqnarray}\label{gpcpl} i \dot \phi_i(x,t) &=& -
\frac {[\phi_i(x,t)]_{xx}}{2} + g_i|\phi_i(x,t)|^2 \phi_i(x,t) \nonumber \\
&+& g_{12}|\phi_j(x,t)|^2 \phi_i(x,t) +
V(x) \phi_i(x,t),
 \end{eqnarray} 
where $i\ne j=1,2$ denote the  
species, and wave functions $\phi_i$
are normalized as
$\int_{-\infty}^{\infty} |\phi_i(x,t)|^2 dx = 1$. 
The suffix $x$ denotes 
space derivatives and overhead dot time derivatives. 
In  (\ref{gpcpl}), time $t$, space 
$x$, nonlinearities $g_i$ and $g_{12}$ are related to the 
 physical observables by
 \cite{MA}: $t= \omega_x t_{\mathrm{phys}}, x=\hat x/l, \{ g_i,g_{12}
\}= 2\{ a, a_{12} \}\lambda /l$ with $l=\sqrt{\hbar/m\omega_x}$ and
$\lambda = \omega_\rho/\omega_x$ is the trap aspect ratio with
$\omega_x$ and $\omega_\rho$ axial and radial ($\rho$) frequencies
and where
 $a_i$ and 
$a_{12}$ are intraspecies and interspecies scattering lengths.
 The 
DW is taken as \cite{DW1} \begin{equation}\label{well1} 
V(x)=x^2/2+Ae^{-\kappa x^2}. \end{equation}
For a 
single-species BEC, the reduced 1D equation
 is 
\begin{eqnarray}\label{gp1c} i \dot \phi(x,t) = -(1/2) \phi_{xx}(x,t)
 + g|\phi(x,t)|^2 
\phi(x,t) + V(x) \phi(x,t). \end{eqnarray}

\begin{figure}
\includegraphics[width=.8\linewidth]{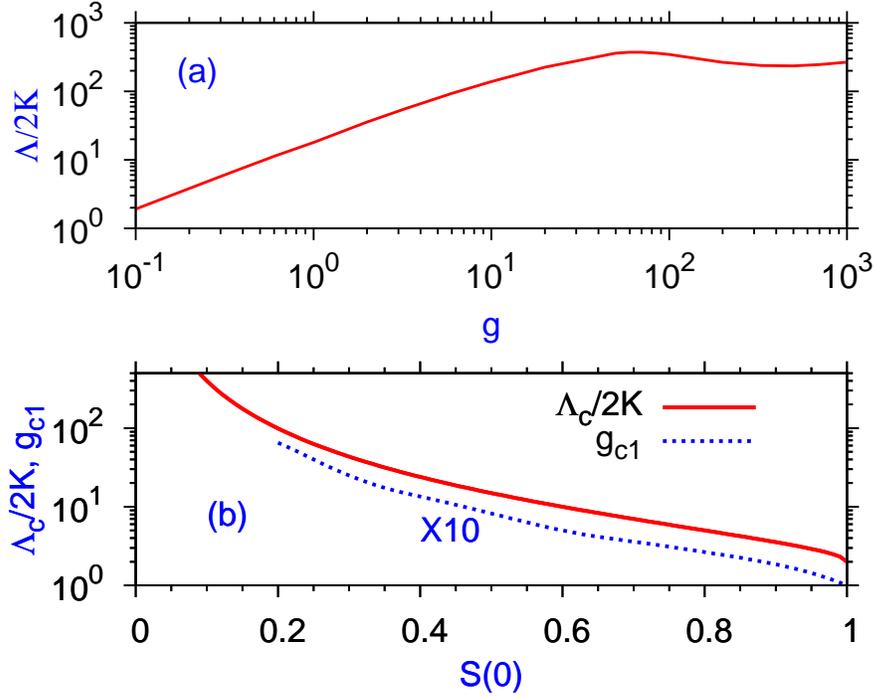}
\caption{ (a) The function   $\Lambda/2K$ versus  $g$
obtained using two-mode functions (\ref{tf}). (b) The critical 
 $\Lambda_c/2K$ and $g_{c1}$ versus $S(0)$ from  (\ref{cn}) and 
(\ref{tf}) for $\theta(0)
=0$. 
}
\label{fig2}
\end{figure}

In the two-mode model \cite{trappingPRL,milburn}  a single-channel BEC 
wave function 
$\phi(x,t)$ of  (\ref{gp1c}) is decomposed as \cite{trappingPRL}
\begin{equation}\label{mmm}
\phi(x,t)=\psi_1(t)\Phi_1(x)+\psi_2(t) \Phi_2(x),
\end{equation}
where spatial modes $\Phi_i(x) (i=1,2)$ in the two wells are 
orthonormalized as $\int\Phi_i(x)\Phi_j(x)=\delta_{ij}$ and 
the functions $\psi_i(t)$ satisfy 
$|\psi_1(t)|^2+|\psi_2(t)|^2=1$. The functions  
$\psi_i(t)$ are  complex and are  separated into its real and 
imaginary parts as $\psi_i(t)=|\psi_i(t)| \exp (i\theta_i)$. 
A population imbalance
\begin{equation}\label{eq7} 
S(t) \equiv (|\psi_1(t)|^2-|\psi_2(t)|^2)/(|\psi_1(t)|^2+|\psi_2(t)|^2)= 
(|\psi_1(t)|^2-|\psi_2(t)|^2)
\end{equation}
 and phase 
difference $\theta \equiv \theta_2-\theta_1$ then serve as a pair of conjugate 
variables.  The approximation (\ref{mmm}) is then 
substituted in  (\ref{gp1c}), and after some straightforward algebra 
we obtain \cite{trappingPRL}
\begin{eqnarray}\label{ds}
\dot S(t)=-2 K \sqrt{1-S^2 (t)} \sin \theta(t),  \\
\dot \theta(t) = \Lambda S(t)+ 2 K\frac{S(t)} {\sqrt{1-S^2 (t)}}\cos \theta(t),
\label{dt}
\end{eqnarray}
where $\Lambda =g\int dx \Phi^4_i(x)$ and 
\begin{eqnarray}
K=-\int dx \Phi_1(x)\left[-\frac{1}{2}\frac{d^2}{dx^2}+V(x)       
  \right]\Phi_2(x).
\end{eqnarray}
The two-mode  equations (\ref{ds}) and (\ref{dt}) are the Hamilton equations 
\cite{trappingPRL}
$\dot S=-{\partial H}/{\partial \theta}, 
\dot \theta=-{\partial H}/{\partial S}, $
for  Hamiltonian 
$
H= {\Lambda}S^2(t)/2  - 2K \sqrt{1-S^2 (t)}\cos \theta(t).$
The transition from Josephson oscillation to self trapping happens at 
 $H=2K$ above a critical $\Lambda = \Lambda_c$ for \cite{trappingPRL}
\begin{eqnarray}\label{cn}
\frac{\Lambda}{2K} > \frac{\Lambda_c}{2K} \equiv \frac{2+2  \sqrt{1-S^2 (0)}\cos 
\theta(0)}{S^2(0)}.
\end{eqnarray}

To perform a numerical calculation using the two-mode model we choose the mode
functions as \cite{trappingPRL}
\begin{equation}\label{tf}
\Phi_{1,2}(x) =[\Phi_+(x)\pm \Phi_-(x)]/\sqrt 2,
\end{equation}
with the property $\Phi_1(-x)=\Phi_2(x)$, 
where $\Phi_\pm(x)$ are the symmetric ground  
and antisymmetric first excited state of  (\ref{gp1c})
with potential (\ref{well1}) with an appropriate $g$. The parameters 
of potential (\ref{well1})
are taken as $A=16$ and $\kappa =10$ (used in most of numerical 
calculations below).   

Now  we calculate the quantity 
$\Lambda/2K$ for different  $g$ and plot in
figure \ref{fig2} (a). Then  a critical $\Lambda_c$ for self trapping 
is obtained from  (\ref{cn}) and plotted in figure  \ref{fig2} (b)
versus $S(0)$ for $\theta(0)=0$. 
From figure \ref{fig2} (b)
we see that 
for $S(0)=0.1$,   the critical $\Lambda_c/2K \sim 400$. From figure \ref{fig2}
(a) we see that this  $\Lambda/2K$ is never attained for any 
nonlinearity $g$. Hence no self trapping can be obtained for $S(0)<0.1$. 
For $S(0)>0.2$, the critical $\Lambda_c/2K$ for self trapping 
as obtained from   figure \ref{fig2} (b), can be attained for $g>g_{1c}$
as seen in figure \ref{fig2} (a). However, with further increase of $g$, 
$\Lambda$ continues to be always greater than $\Lambda_c$ and the self trapping 
is never destroyed with the increase of $g$. In our numerical study we shall 
find that the self trapping is destroyed with the increase of $g$. 
This is reasonable as the 
two-mode model with the neglect of overlap integrals between the mode 
functions $\Phi_{1,2}(x)$
is expected to be valid for 
small  $g$ only. 
A critical nonlinearity for self trapping 
$g_{1c}$ for $\theta(0)=0$ and  different $S(0)$ is then obtained from 
the results of $\Lambda/2K-g$ and  $\Lambda_c/2K-S(0)$ plots 
in figures  \ref{fig2}  (a) and (b) and 
 is also plotted in 
figure \ref{fig2} (b).

\section{Numerical results for Double-Well (DW) potential}

\label{III}

The GP equations  are
solved numerically using the Crank-Nicolson
scheme \cite{MA,MA2}
with space and time steps 0.025 and 0.0002,
respectively, by real-time propagation with  FORTRAN programs
provided in  \cite{MA}.

\subsection{Single-Channel BEC}

Here we consider DW  (\ref{well1}) with  $A=16$
and $ \kappa=10$, as in  some previous studies \cite{DW1,hanpu} on 
self trapping. 
In numerical simulation
we create an initial state with a fixed  population imbalance $S(0)$, 
by looking for the ground state at $x=x_0$ of 
the following asymmetric well  \cite{DW1}
\begin{eqnarray}\label{asy}
U(x)=(x-x_0)^2/2+Ae^{-\kappa x^2} .  
\end{eqnarray}
The initial $S(0)$ is achieved by 
varying the  parameter
$x_0$ in (\ref{asy})  \cite{DW1}. 
Once this initial state is created,  $x_0$ is reduced to zero so that 
DW (\ref{asy}) reduces to DW (\ref{well1})
with $x_0=0$. If $x_0$ 
is reduced  slowly during  time evolution, 
the initial asymmetric state relaxes to a symmetric stationary state 
of  DW (\ref{well1}) and there is no self trapping. 
On the other hand if $x_0$ is reduced  
quickly, there is always self trapping provided that  
$g$ lies in the appropriate window: $g_{1c}<g<g_{2c}$.  
This is shown in figure \ref{fig3}, 
where we plot the density at different $t$ after
$x_0$ is reduced to zero during an interval 
$\Delta t= 5, $
and 20. 
From figure \ref{fig3}  
we find that there is self trapping for $\Delta t=5$  and no self 
trapping for $\Delta t =20$, when the asymmetric initial state 
relaxes to a final symmetric state. 
 By actual substitution 
of the wave function of the self-trapped state
in the {\it time-independent} GP equation we verify that 
there is no stationary state of the DW with this density. 
The present trapped wave function was generated from the ground
state of DW (\ref{asy}) and is always positive with no node and the 
present self trapping is dynamical  without any static counterpart.

\begin{figure}
\includegraphics[width=.8\linewidth]{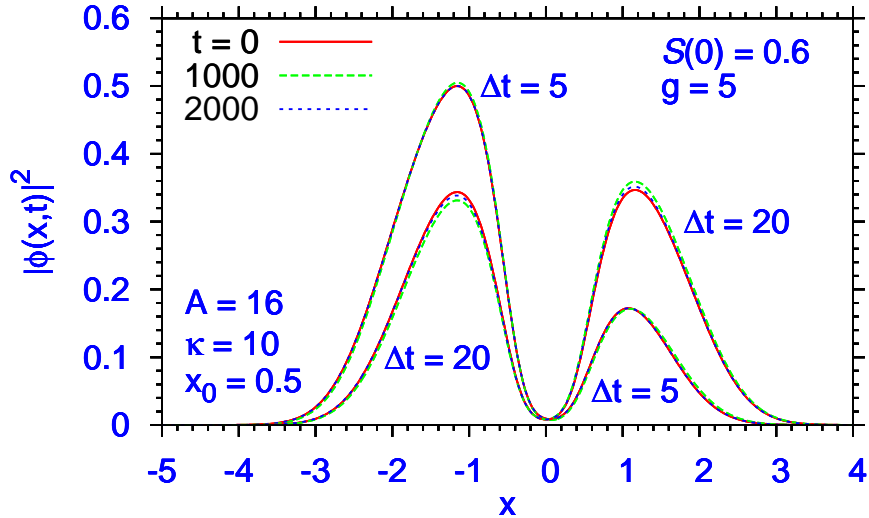}
\caption{Density $|\phi(x,t)|^2$ versus $x$ at different $t$ when an
initial state,
created as the
ground state of the asymmetric DW  
(\ref{asy}), is allowed to evolve in time while the  asymmetric    DW
(\ref{asy})
 is changed to the symmetric DW  (\ref{well1}) 
by reducing $x_0$ to
zero slowly in an interval of time    $\Delta t=5,20.$ When the transition 
is quick ($\Delta t = 5$)  a permanent self trapping with an asymmetric 
profile emerges. For a slow transition ($\Delta t=20$), the system smoothly 
moves to the symmetric ground state of the symmetric DW  (\ref{well1}) and 
there is no self trapping. }
\label{fig3}
\end{figure}

The self trapping is  sensitive to  $S(0)$. 
There is no self trapping for a small  
$S(0) (< 0.1)$ (viz., two-mode model of section \ref{IIB}).
In figure \ref{fig4} (a) (upper panel)
we plot  $S(t)$ versus $t$  to illustrate the 
trapping and 
oscillation for different  $g$ $(=1,10,100,1000)$ for  $S(0)=0.3$. 
  For   $g=1$, there is
Josephson oscillation with $\langle S(t)\rangle=0. $ For $g=10$ we
 have $\langle S(t)\rangle \sim 0.3$  insuring self 
trapping. As $g$ is 
further increased, the nonlinear term in the GP equation becomes much 
larger than the Gaussian wall $A\exp(-\kappa x^2)$ in  (\ref{well1}). 
Consequently, {the DW appears 
like a single well and classical center-of-mass oscillation 
appears  for a large 
$g$. For $g=100$ the  oscillation is irregular because the 
Gaussian wall  is not fully negligible. Finally, for $g=1000$ 
 the Gaussian wall  can be completely neglected 
and we have clean classical center-of-mass oscillation.}
The $S(t)$ versus $t$ plot for $S(0)=0.3$ and $g=10$ at large 
times is shown in the lower panel of figure \ref{fig4} (a), where 
self trapping is confirmed up to $t=5000$ $-$ an interval much  
larger than tunneling time which is at best $\sim 100$.
The evolution of $S(t)$ at 
large times $S(\infty)$ for 
different $S(0)$ as $g$ is increased is shown in figure \ref{fig4}
(b). A nonzero $ S(\infty)$ ensures self trapping,  which appears  
 in the window: $g_{c1}<g<g_{c2}$. 
The prediction of the analytical two-mode model of section \ref{IIB}, from 
figure 
\ref{fig2} (b), 
for $g_{c1}$ are shown by arrows in figure \ref{fig4} (b) for $S(0)=0.3$ 
and 0.6  in good agreement with numerical 
simulation. { It is noted that in all calculations reported 
here $g_{c2}<100$
$-$ a medium value of nonlinearity where the mean-field GP description 
is well justified. For larger values of nonlinearities, beyond 
mean-field corrections  
to the GP equation \cite{skls}
could be relevant 
\cite{bmf}. However, we expect 
the general conclusions of this study  to remain valid for the present 
set of parameters.}

\begin{figure}
\includegraphics[width=.8\linewidth]{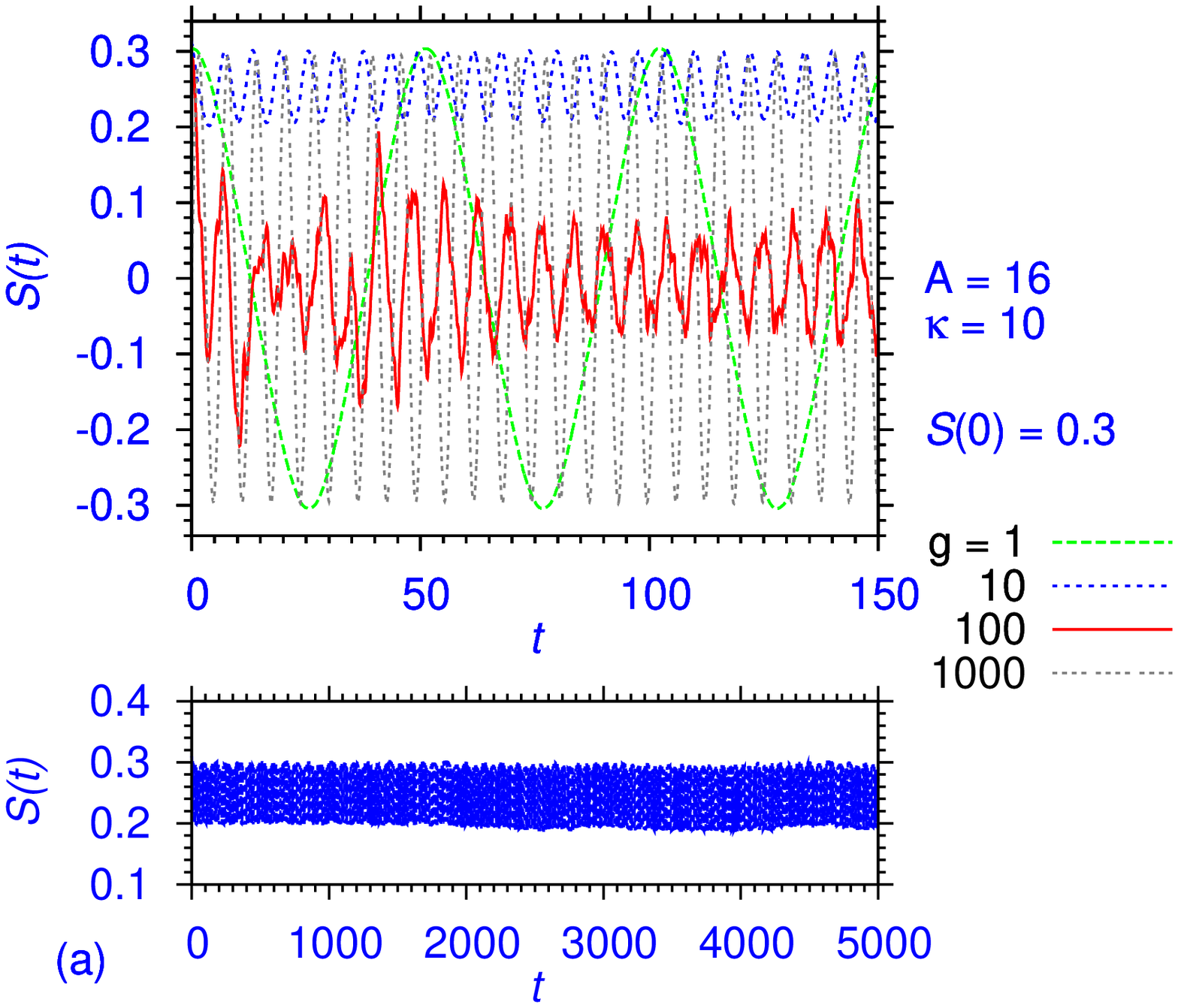}
\includegraphics[width=.8\linewidth]{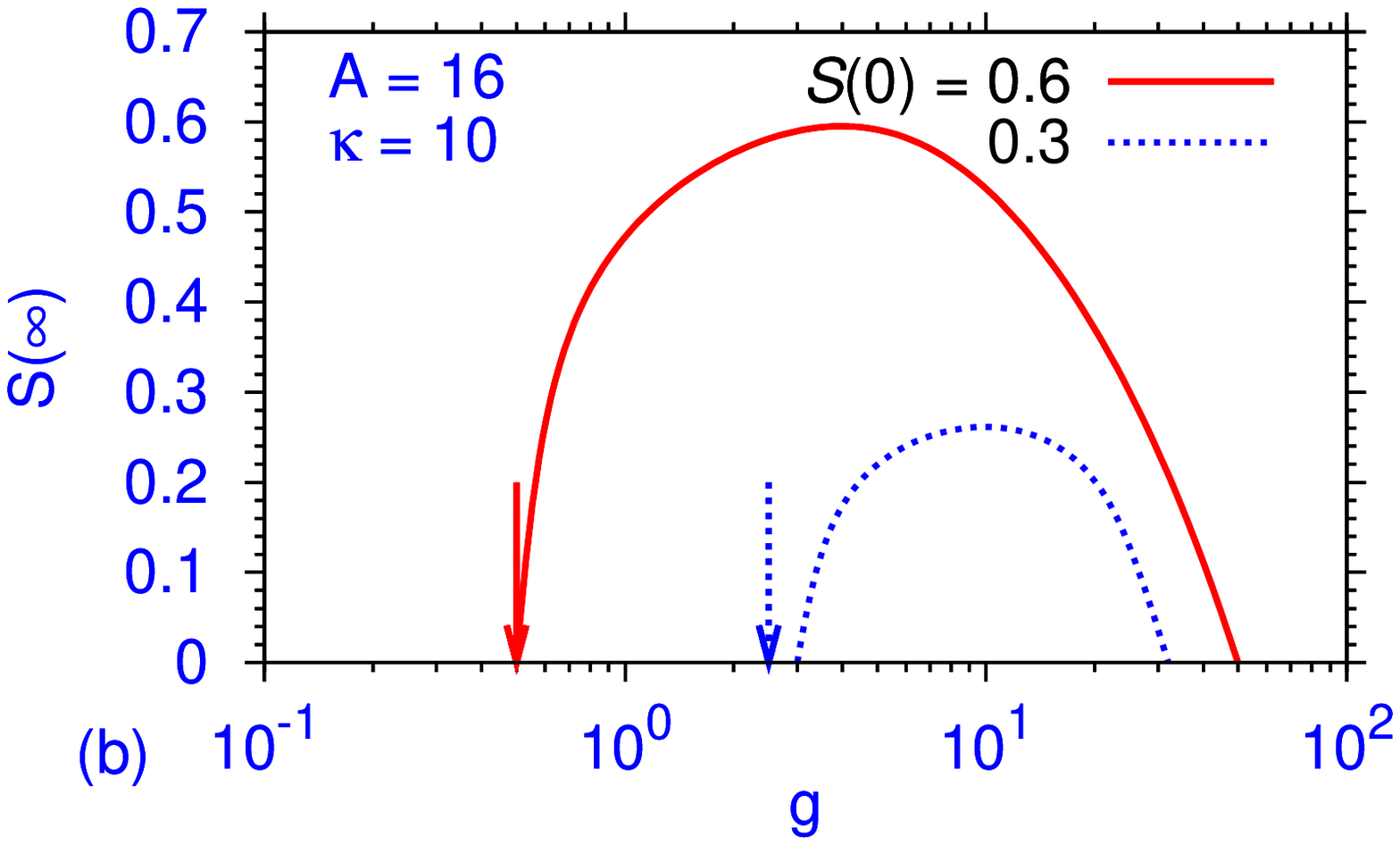}
\caption{(a) (upper panel) Population imbalance $S(t)$ versus $t$ 
for  $S(0)=0.3$
 in DW
(\ref{well1}).
(lower panel) Long-time $S(t)$ versus $t$  for $g=10$.
(b) Average population imbalance at large time 
$S(\infty) $ versus 
 $g$ for $S(0)=0.3$ and 0.6. The arrows are  
results for   $g_{c1}$ 
from figure \ref{fig2} (b).}
\label{fig4}
\end{figure}

\begin{figure}
\includegraphics[width=.8\linewidth]{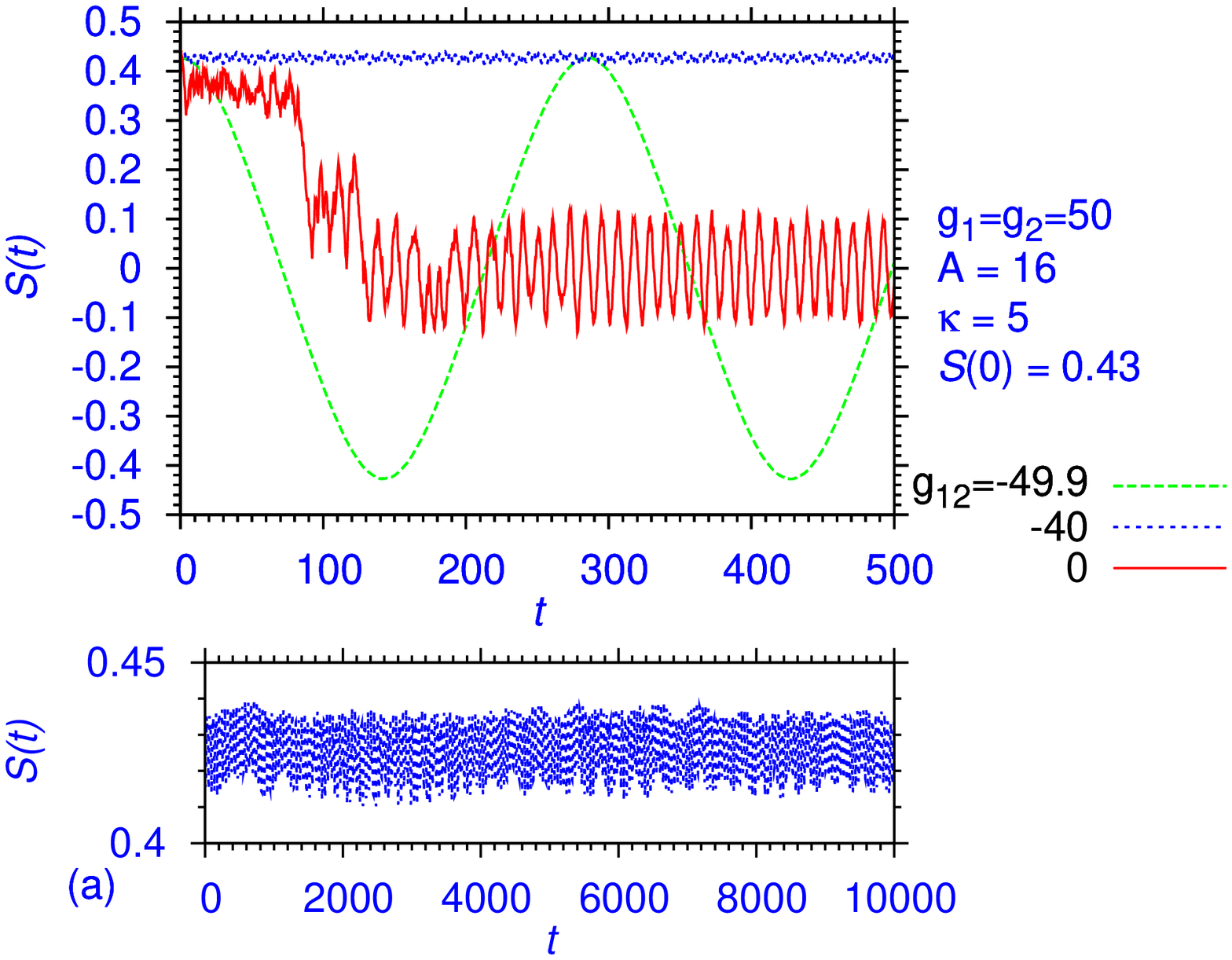}
\includegraphics[width=.8\linewidth]{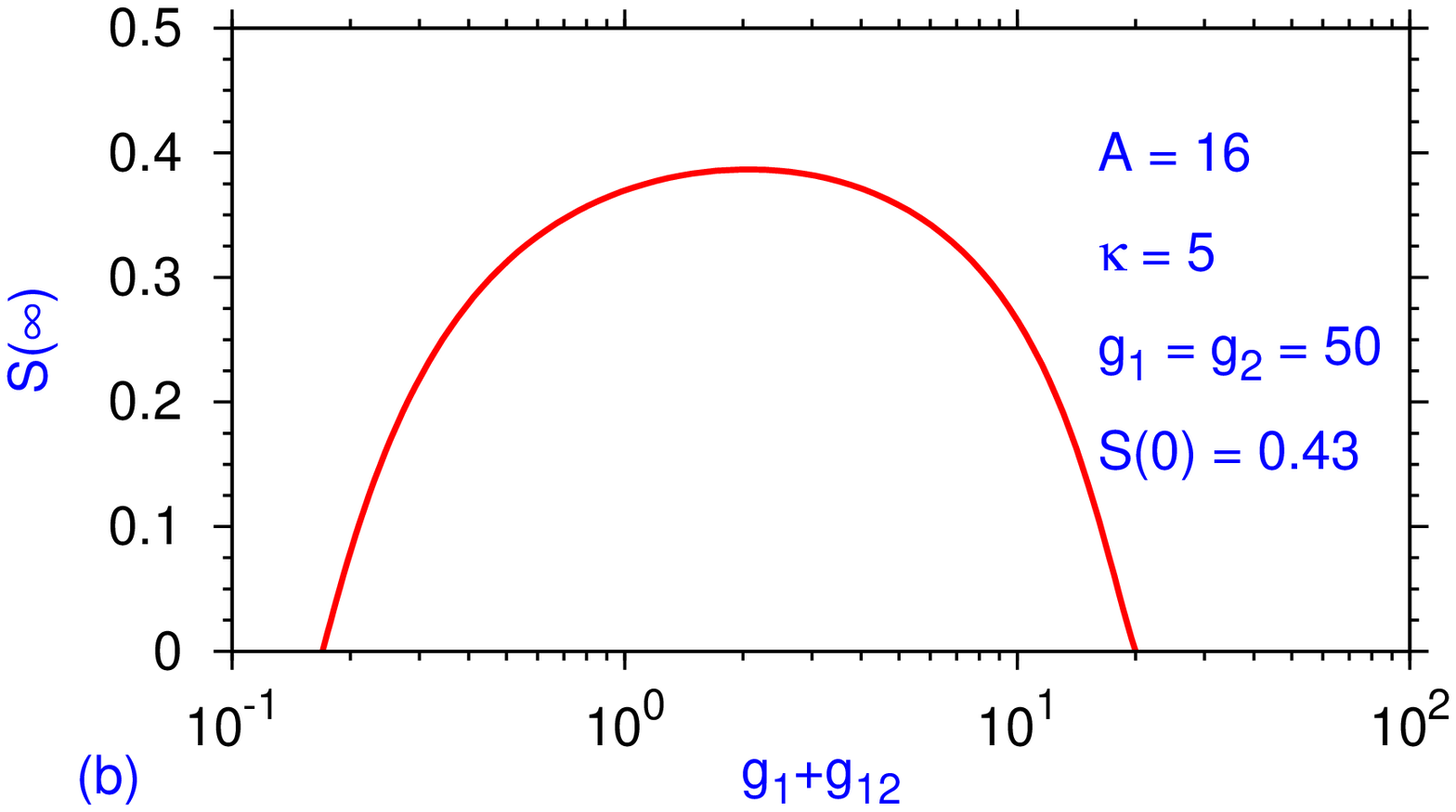}
\caption{(a) (upper panel) Population imbalance $S(t)$ versus 
$t$ 
for a binary BEC with $S(0)=0.43$ and 
nonlinearities $g_1=g_2=50$ 
and    $g_{12}=0,-40$ and $-49.9$
 with  DW  (\ref{well1}).
(lower panel) Long-time $S(t)$ versus $t$ dynamics for $g_{12}=-40$.
(b) Average population imbalance $S(\infty) $ at large time versus 
effective nonlinearity $g_1+g_{12}$ for $S(0)=0.43$.  }
\label{fig5}
\end{figure}

\subsection{Binary BEC}

We consider the self trapping of a binary BEC in a DW 
for  $g_i>g_{c2}$ and $g_i<g_{c1}$. In both 
cases no self trapping is possible for zero interspecies 
interaction ($g_{12}=0$). Of these, for $g_i<g_{c1}$ we need a 
repulsive (positive) $g_{12}$ to make both components sufficiently 
repulsive to have self trapping. But this case usually 
leads to a uninteresting stationary phase separated configuration \cite{PS}
with the first species 
occupying the first well of the DW and the second species 
occupying the second well. 
 A more interesting situation emerges for $g_i>g_{c2}$, 
when we require an attractive (negative) $g_{12}$ to make both 
components appropriately repulsive to have self trapping.
However, self trapping appears only for an intermediate $S(0)$ 
for  an appropriate  $g_i$ and $g_{12}$. 
To illustrate, 
we consider the symmetric case with $g_1=g_2=50$ in  
(\ref{gpcpl})  and (\ref{well1}) and consider an attractive (negative) 
 $g_{12}$. In the binary case, the self trapping 
is not so good for $\kappa=10$, while the barrier between the two wells in 
 (\ref{well1}) is narrow and we take $\kappa=5$ in the numerical 
simulation.  
Permanent self trapping is found to occur 
for a narrow window of $S(0)$  around 
$S(0)=0.4$. For $S(0)>0.5$ and $S(0)<0.3$ self trapping for a small 
interval of time could be obtained. 

The initial state with 
$S(0)=0.43$ for both 
components
is created using DW
 (\ref{asy}) and  reducing  $x_0$ 
to zero.  
In the upper panel of figure \ref{fig5} (a), we 
plot  $S(t)$ versus $t$ for both 
components for three values of $g_{12}$. For $g_{12}\ge 0$, no self trapping 
is possible and we are in the domain of irregular oscillation as 
shown for $g_{12}=0$. For an  attractive  $g_{12}$, 
there is permanent self trapping as illustrated for $g_{12}=-40$. With further 
increase of $|g_{12}|$ the self trapping disappears and regular sinusoidal 
Josephson oscillation appears as shown for $g_{12}=-49.9$ in figure \ref{fig5} 
(a).
The lower panel
 of figure \ref{fig5} (a) illustrates 
the 
population imbalance $S(t)$ for $g_{12}=-40$
at large 
times,  which confirms  robust self trapping. 
In figure \ref{fig5} (b) we plot  $S(\infty)$ for both components versus 
$g_1+g_{12}$. In the symmetric case with $g_1=g_2$, the wave function 
of the two components are equal and the quantity $g_1+g_{12}$ is the 
effective nonlinearity of each component. Figure \ref{fig5} (b) shows that 
self trapping occurs for the window of effective nonlinearity 
$0.2<g_1+g_{12}<20$ for $S(0)=0.43$.
{ 
For any given initial 
population imbalance and for either sufficiently small or sufficiently 
large effective nonlinear interaction strength
$g_1+g_{12}$, the system is in the 
oscillation regime. For small values of  effective interaction one 
has Josephson oscillation and for large values of effective interaction
one has the classical center-of-mass oscillation.} 
For intermediate interaction strength, the system 
may make transition to self trapping for an appropriate  $S(0)$.

\section{Analytical formulation for Optical-Lattice (OL) potential}

\label{IIA}

For an OL, Eqs. (\ref{gpcpl}) and (\ref{gp1c}) remain valid but with the 
variables bearing the following relations to the corresponding physical 
observables
\cite{am}: $t=(\pi/L)^2
(\hbar/m)t_{\mathrm{phys}}, x=\hat x\pi/L,
V_0=m(L/\pi\hbar)^2 V_{ol},
\{g_i,g_{12} \}=(2NLm\omega_\perp /\pi \hbar)
\{a_i,a_{12} \},$ where   $L$ the wavelength and
$V_{0}$ the strength of OL potential: $V_{OL}(\hat x)=-V_{0}
\pi^2 \hbar^2/(mL^2) \cos (2\pi\hat  x /L)$. The dimensionless OL potential 
is given by 
\begin{equation}
V(x)= -V_0\cos(2x).  \label{olpot}
\end{equation}

\begin{figure}
\includegraphics[width=.49\linewidth]{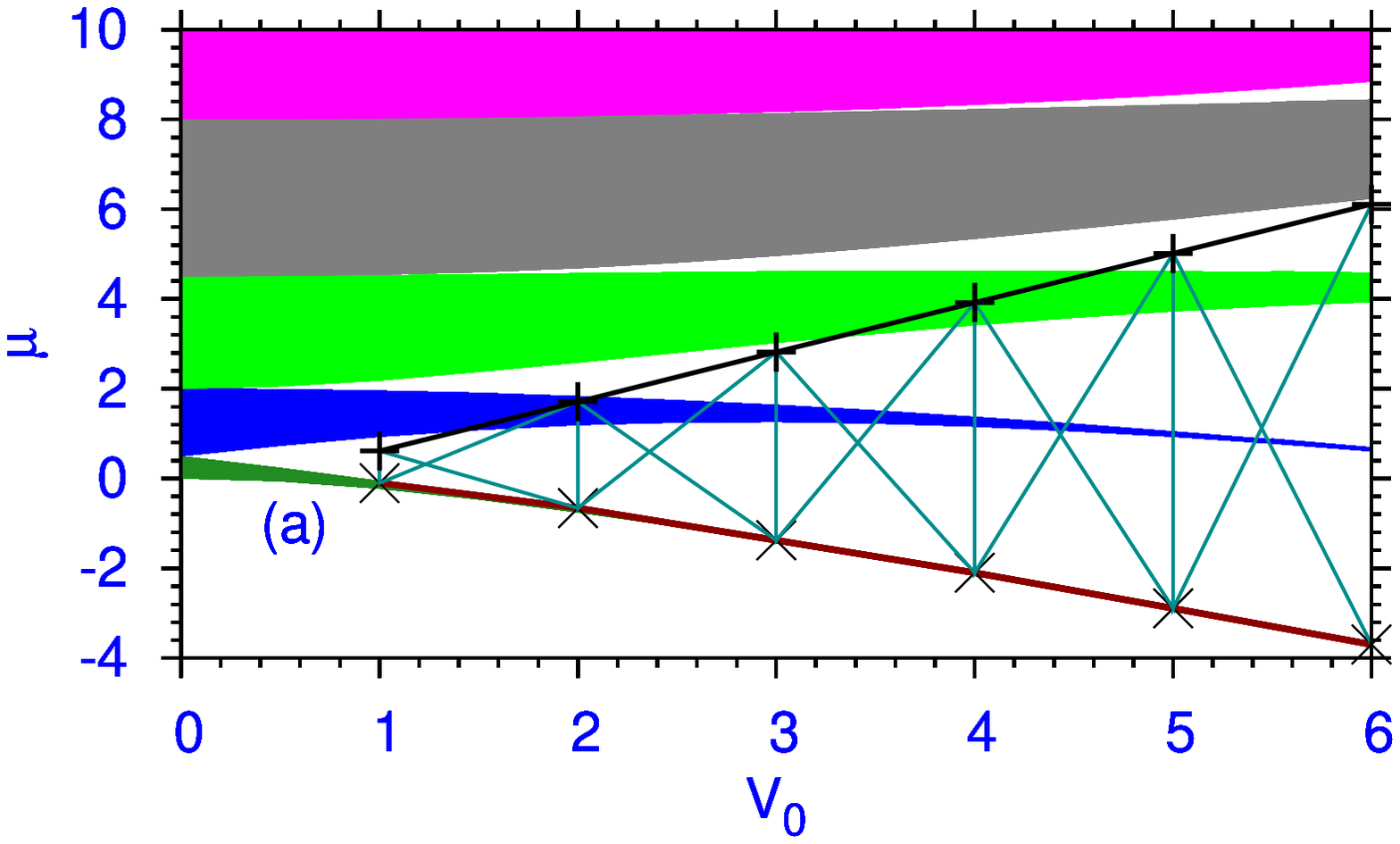}
\includegraphics[width=.49\linewidth]{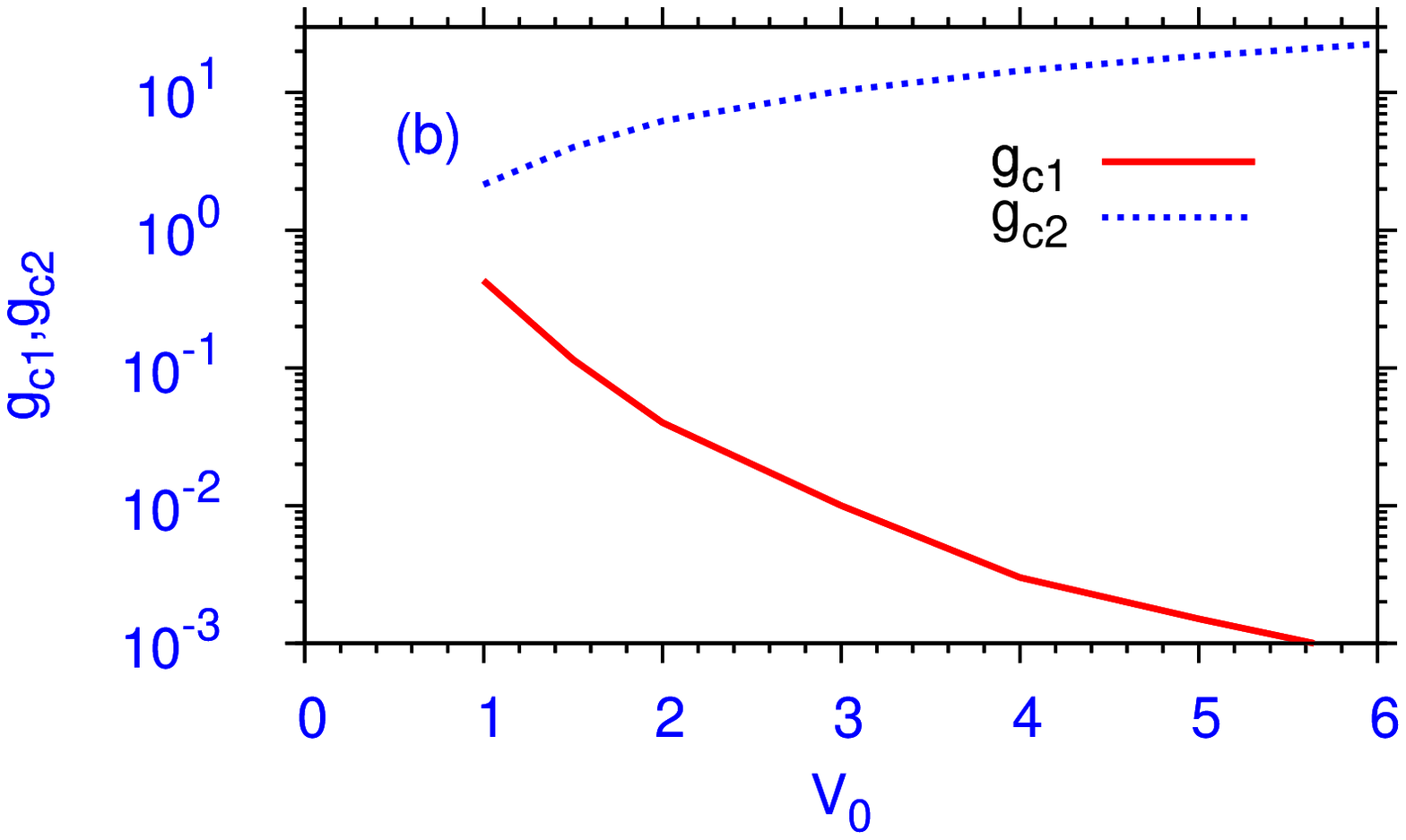}
\caption{(a)  
Chemical potential $\mu$ versus  $V_0$
showing the bands (shaded area)
and gaps (white area) for OL (\ref{olpot}).
{
The hatched area between  
the lower line with crosses and the upper line with pluses
represents the {\it variational} domain of the appearance of Gaussian gap solitons.
The full white area above the lower line with crosses denote the {\it exact} 
domain of the appearance of gap solitons.}
(b) The critical values $g_{c1}$ and $g_{c2}$ defining the window
of nonlinearities of Gaussian gap solitons 
versus $V_0$. 
}
\label{fig1}
\end{figure}

To critically examine the self trapping in a 
OL where the BEC is  localized in one of the sites 
of the OL, we present  a Gaussian variational analysis. The stationary 
state is described by  (\ref{gp1c}) with the time derivative term 
$i\dot \phi$ replaced by $\mu \phi$ where $\mu$ is the chemical potential. 
The Lagrangian for that stationary equation is given by  \cite{gap2} 
\begin{eqnarray}\label{lag}
L=\int_{-\infty}^\infty\left[ \mu \phi^2(x)-\phi_x^2(x)/2-g\phi^4(x)/2+
V_0\cos(2x)\phi^2(x)\right]dx-\mu .
\end{eqnarray}
{This Lagrangian can be analytically evaluated by considering a simple 
form for the wave function $\phi$. Using the  Gaussian form 
 \cite{vari}}
$
\phi(x)=\pi^{-1/4}\sqrt{{\cal N}/w}\exp[-x^2/(2w^2)]
$
with ${\cal N}$ the norm and $w$ the width, the Lagrangian can be written
as 
\begin{eqnarray}
L=\mu({\cal N}-1)-\frac{\cal N}{4w^2}+V_0{\cal N}\exp(-w^2)
-\frac{g{\cal N}^2}{2\sqrt{2\pi}w}.
\end{eqnarray}
The variational equations \cite{vari}
$\partial L/\partial \mu= \partial L/\partial w
=\partial L/\partial {\cal N}=0$ yield ${\cal N}=1$ and 
\begin{eqnarray}\label{wi}
1&+&\frac{gw}{\sqrt{2\pi}}=4V_0w^4\exp(-w^2), \\  \label{mu}
\mu&=& \frac{1}{4w^2}+\frac{g}{\sqrt{2\pi }w}- V_0\exp(-w^2),
\end{eqnarray}
which determine the width and the chemical potential. 
{We have set ${\cal N}=1$
in  (\ref{wi}) and (\ref{mu}) after derivation.
}

\begin{figure}
\includegraphics[width=.8\linewidth]{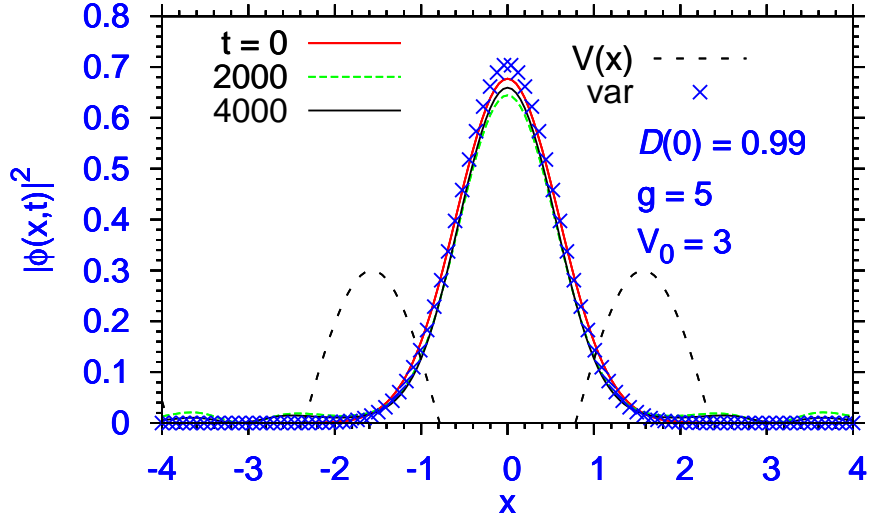}
\caption{
Density $|\phi(x,t)|^2$ versus $x$ at different $t$ when an
initial Gaussian 
state with trapping measure  $D(0)=0.99$ is released at $x=t=0$ on
the OL. The variational result for the corresponding 
gap soliton is shown by crosses.  
The positive part of the potential  is also 
shown in arbitrary units. 
}
\label{fig6}
\end{figure}

{It is instructive to study the band and gap structure of the 
Schr\"odinger equation in the periodic OL potential (\ref{olpot}): $\mu 
\phi(x)= -(1/2)\phi _{xx}(x)-V_0\cos(2x)\phi(x)$ \cite{kittel}. In a periodic 
potential the quantum excitation spectrum of a system consists of bands 
and gaps. The bands allow unlocalized plane wave solution modulated by a 
periodic function with the same period as the periodic potential known 
as Bloch wave. The gaps permit localized solution \cite{kittel}. 
However, the single-particle linear Schr\"odinger equation does not 
allow any solution in the gap. The band (shaded regions) and gap (white 
regions above the lowest shaded region) of the spectrum of the 
Schr\"odinger equation with OL potential (\ref{olpot}) is shown in 
figure \ref{fig1} (a). The nonlinear 1D GP equation (\ref{gp1c}) for 
repulsive interaction (positive $g$) permits localized 
solutions in the gaps, called {\it gap solitons}, 
where the chemical potential $\mu$ lies in the gap.  In the gaps, 
localized gap solitons are possible in the presence of 
an appropriate nonlinearity $g$.}

We now find the condition for a gap soliton for a positive (repulsive) 
$g$ by directly solving (\ref{wi}) and (\ref{mu}) for all $g$ and $V_0$. 
As $\mu $ tends to the upper edge of the lowest band, one has the lower 
limit $g_{c1}$
of the formation of a gap soliton, denoted by a line with crosses 
in figure \ref{fig1} (a). As the repulsive nonlinearity $g$ is 
increased, (\ref{wi}) and (\ref{mu}) permit solution up to a maximum 
value $g_{c2}$ 
which determine the upper limit of the formation of a gap soliton 
denoted by the line with pluses in figure \ref{fig1} (a). 
{The area 
between the two lines determine the domain in which the Gaussian gap 
solutions are allowed. (Gap solitons of non-Gaussian shape, possibly occupying 
many OL sites, are possible in the whole white region above the lowest 
band gap in figure \ref{fig1} (a).) }
The nonlinearities corresponding to these two lines 
can be calculated using (\ref{wi}) and (\ref{mu}) and yields the 
critical $g_{c1}$ and $g_{c2}$ for self trapping.  These nonlinearities 
are plotted in figure \ref{fig1} (b).

\section{Numerical results for Optical-Lattice (OL) potential}

\subsection{Single-Channel BEC}

The numerical simulation is started by releasing a 
Gaussian state in the center of the OL during 
time evolution of the GP equation. The width 
of the Gaussian is taken such that the initial state stays
mostly in the central well of the OL. An estimate  of  self trapping is 
given by the following function called trapping measure
\begin{equation}\label{dyn}
D(t)= \int_{-\pi/2}^{\pi/2} |\phi (x,t)|^2 dx.
\end{equation}
Here we note that $\pi$ is the wave length of the OL  $-V_0\cos(2x)$. 
Hence the function $D(t)$ determines the matter inside a single OL site. 
In case of ideal self trapping  in a single site $D(t)=1$, and $D(t)$ will
tend to   zero when self trapping is fully destroyed.

Self trapping occurs easily if the nonlinearity 
 $g$ is appropriate for a Gaussian gap soliton  (viz. figure 
\ref{fig1}). We take an initial Gaussian
state with  $D(0)= 0.99$ and release it on the OL  
with $V_0=3$ and $g=5$.  The  
density profile of the trapped state are shown in figure 
\ref{fig6}. The small dispersion of the density profile at different $t$
guarantees good 
self 
trapping. 
We also plot the density of the 
variational gap soliton, in good agreement with the self-trapped state, 
in figure \ref{fig6}.

\begin{figure}
\includegraphics[width=.49\linewidth]{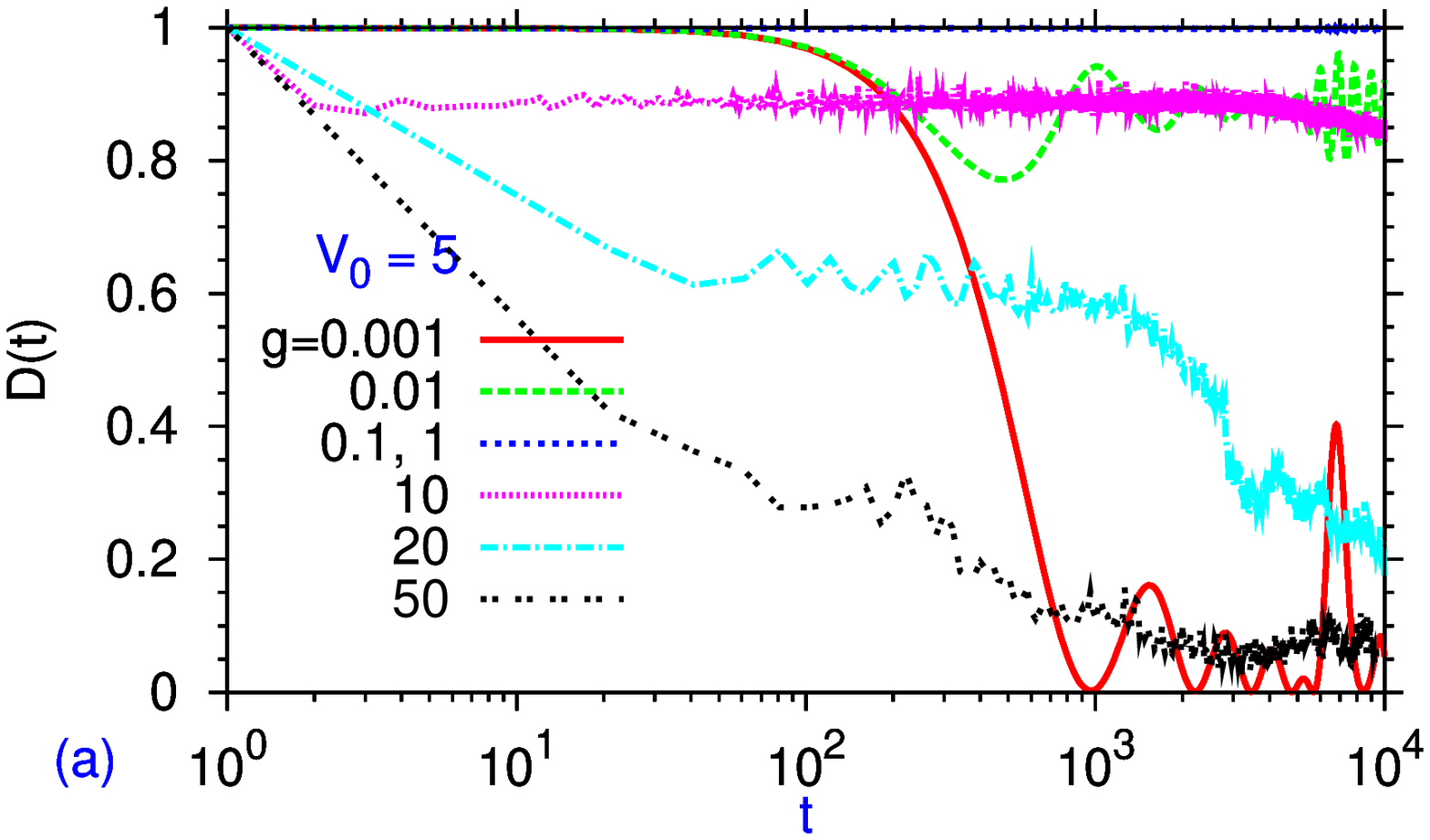}
\includegraphics[width=.49\linewidth]{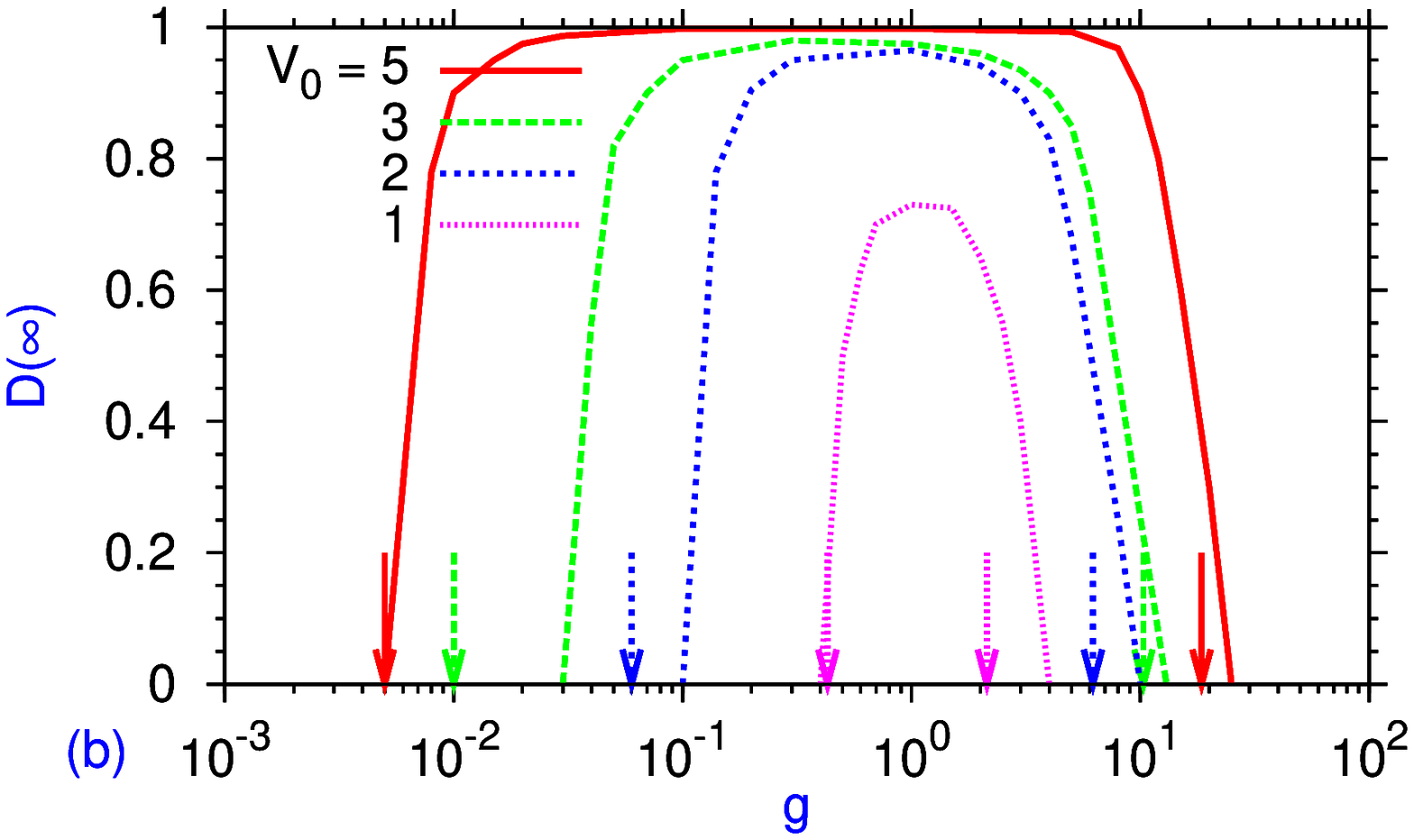}
\caption{
(a) Trapping measure $D(t)$ versus time $t$ for $V_0=5$ 
for different $g$. 
(b) Trapping  measure at large time $D(\infty)$ versus $g$ for 
 different $V_0=1,2,3,5$. 
The respective arrows represent  
nonlinearities $g_{c1}$ and $g_{c2}$ from 
figure \ref{fig1} (b).
}
\label{fig7}
\end{figure}

In Fig \ref{fig7} (a) we 
plot  $D(t)$ versus $t$ for $V_0=5$ and different $g$. 
For $g <10$, $D(t)$ remains close to 1 for  $t<100$. 
However, if we continue to large $t (<10000)$ (a time much larger than 
the tunneling time of few hundred as in Fig \ref{fig7} (a)), 
$D(t)$ remains 
close to 1 for a window of nonlinearity $0.01<g<10$ denoting
permanent self trapping.   For illustration,
in figure \ref{fig7} (b), we plot 
the  trapping measure $D(t)$ at large 
times $D(\infty)$ versus $g$, where  
 $D(\infty)$ is non-zero in the window 
$g_{c1}<g<g_{c2}$ corresponding to permanent self trapping 
and is zero outside showing no
self trapping. 
In figure \ref{fig7} (b) there are two arrows for each 
$V_0$ corresponding to $g_{c1}$ and $g_{c2}$ as obtained in 
 figure \ref{fig1} (b). The domain between the two arrows representing 
the region of allowed gap solitons (see, figure \ref{fig1}) agrees well 
with the domain of self trapping represented by non-zero $D(\infty)$ as 
seen in figure \ref{fig7} (b). From this fact and also from the 
proximity of the variational solution for density of a gap soliton with 
that for the numerical self-trapped state in figure \ref{fig6}, we 
conclude that the self trapped state represents small oscillation of a 
stable gap soliton.  {In this connection, the finding in Ref.
\cite{ol1}, that a self-trapped state in an OL is always temporary, is not
fully to the point. They missed the fact that for the window of nonlinearity 
$g_{c1}<g<g_{c2}$ the self trapping could occur in a stable 
stationary gap soliton state leading to a permanent trapping. Just outside this 
small window of nonlinearity a temporary self trapping at small times 
may take place, as can be seen in figure \ref{fig7} (a).}

\begin{figure}
\includegraphics[width=.49\linewidth]{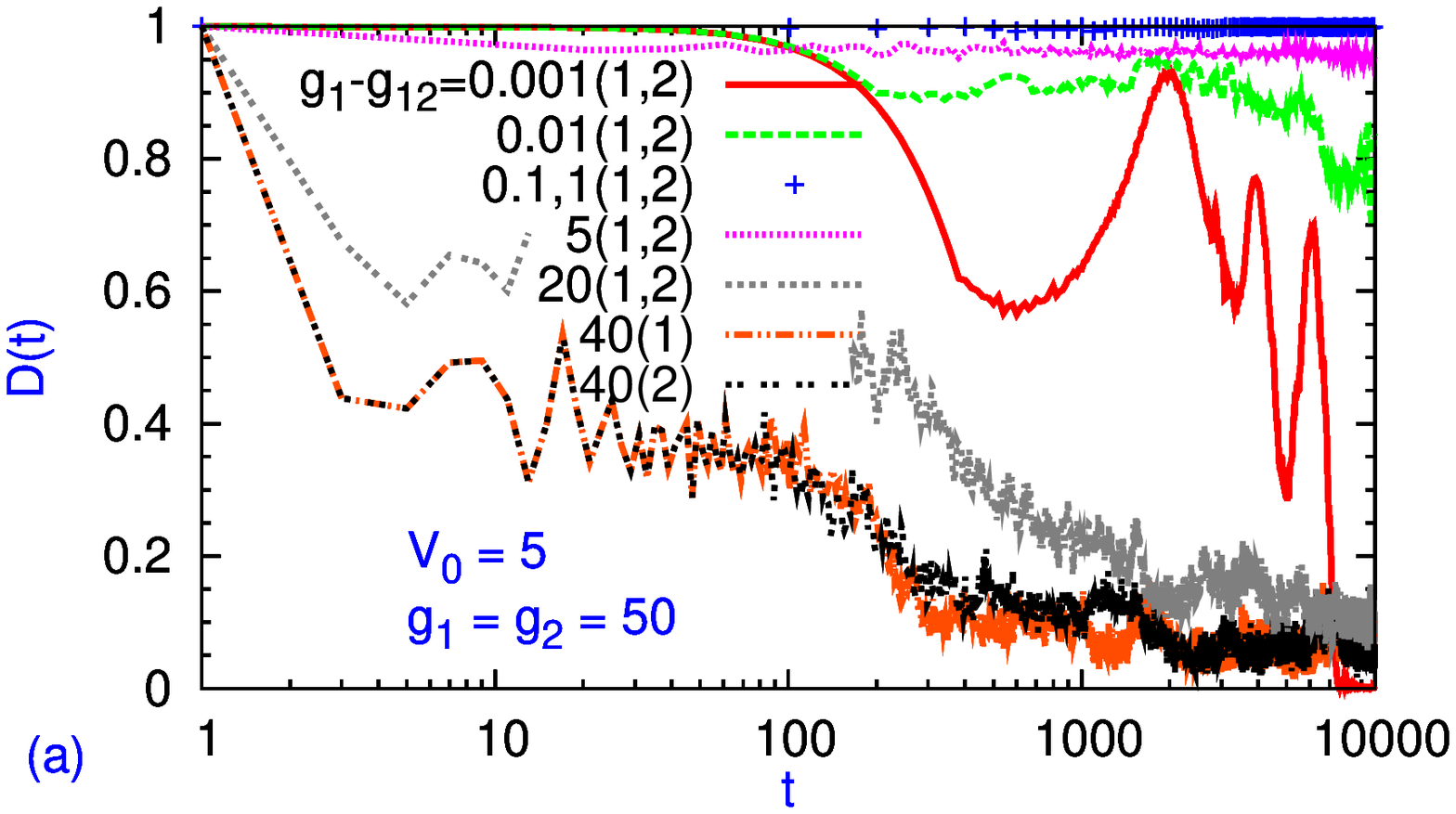}
\includegraphics[width=.49\linewidth]{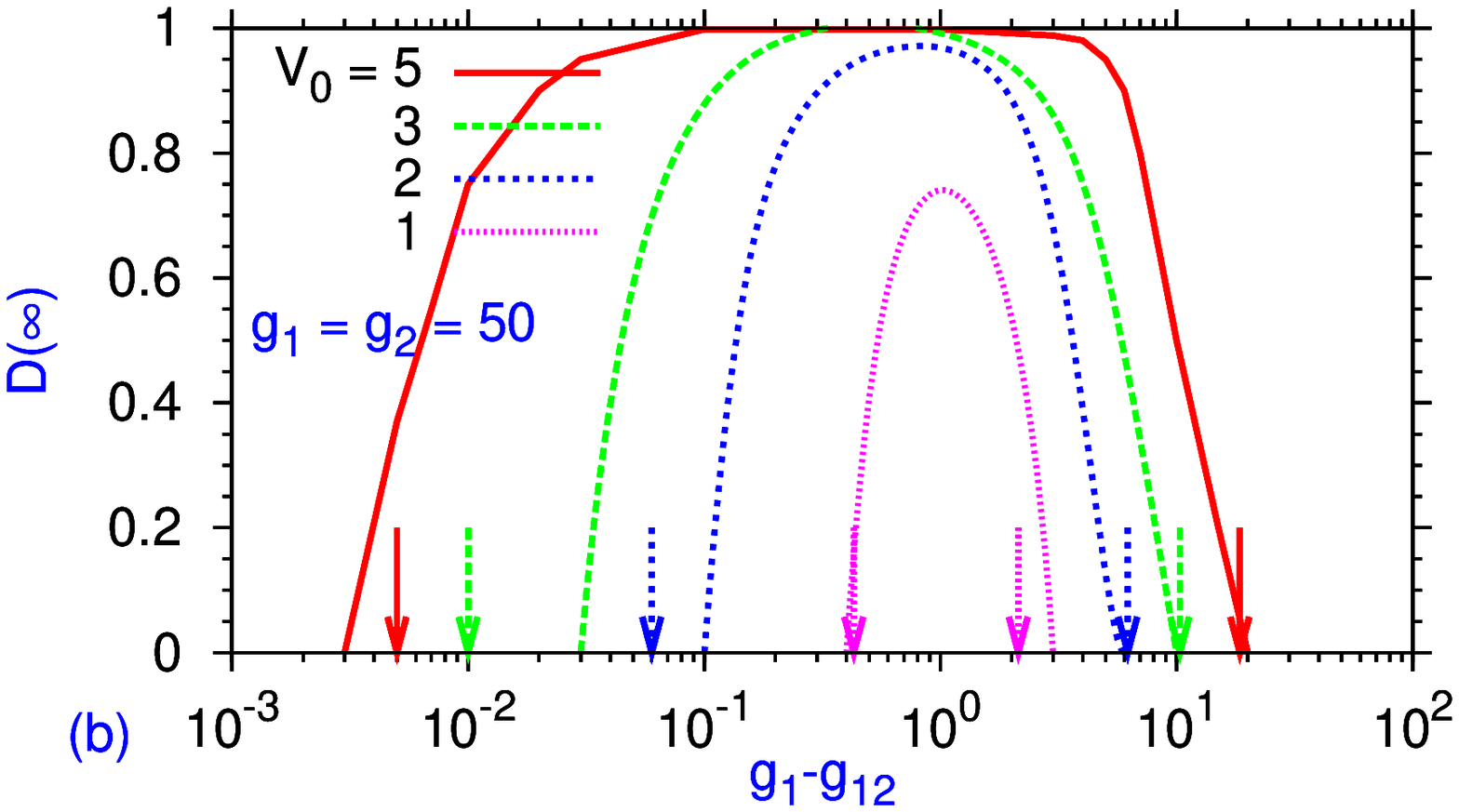}
\caption{
(a) Trapping measure $D(t)$ of a binary BEC 
versus time $t$
for different $g_1-g_{12}$. The curves are labeled by  
$g_1-g_{12}$ and 
the component number in parenthesis.
(b) Trapping  measure at large time $D(\infty)$ versus $g_1-g_{12}$  
for different $V_0$.  
The arrows are 
$g_{c1}$ and $g_{c2}$  from 
figure \ref{fig1} (b).
}
\label{fig8}
\end{figure}

\subsection{Binary BEC}

To study self trapping of a binary BEC 
we consider  large repulsive intraspecies nonlinearities
$g_1=g_2 =50,$ which do not 
allow self trapping for zero  interspecies interaction 
$g_{12}=0$ for $V_0=1,2,3,5,$ as seen in  figures  \ref{fig7} (a) and 
(b). If we introduce an attractive (negative) interspecies 
interaction $g_{12}$, then in each channel the effective nonlinearity 
will be reduced and for a sufficiently large and attractive $g_{12}$ one 
can have self trapping. The trapping dynamics for this system is 
illustrated in figure \ref{fig8} (a) where we plot trapping measure 
$D(t)$ versus $t$ for different $g_{12}$ and $V_0=5, g_1=g_2 =50$. The results for 
the two components are practically the same in most cases.
The self trapping appears for a small $(g_1+g_{12}) (\approx 0.01)$
and disappears for large $(g_1+g_{12}) (> 10) $. In this symmetric binary BEC,
$(g_1+g_{12})$ provides a good measure of the effective nonlinearity 
 controlling self trapping. In figure \ref{fig8} (b) we plot the trapping 
measure at large times $D(\infty)$ versus effective nonlinearity $g+g_{12}$ 
of the binary BEC for $V_0=1,2,3,5$. The plots of figures \ref{fig7} (b) 
and \ref{fig8} (b) are qualitatively quite similar, showing
that $(g_1+g_{12})$ is a good measure of effective nonlinearity of the
binary BEC. However, if $|g_{12}|$ is taken to be larger than $g$ then 
the effective nonlinearity becomes attractive corresponding to a negative 
$g_1+g_{12}$. This domain of nonlinearity corresponds to permanent 
symbiotic 
bright soliton \cite{brightcpl}
and consideration of self trapping is inappropriate.

\begin{figure}
\includegraphics[width=.49\linewidth]{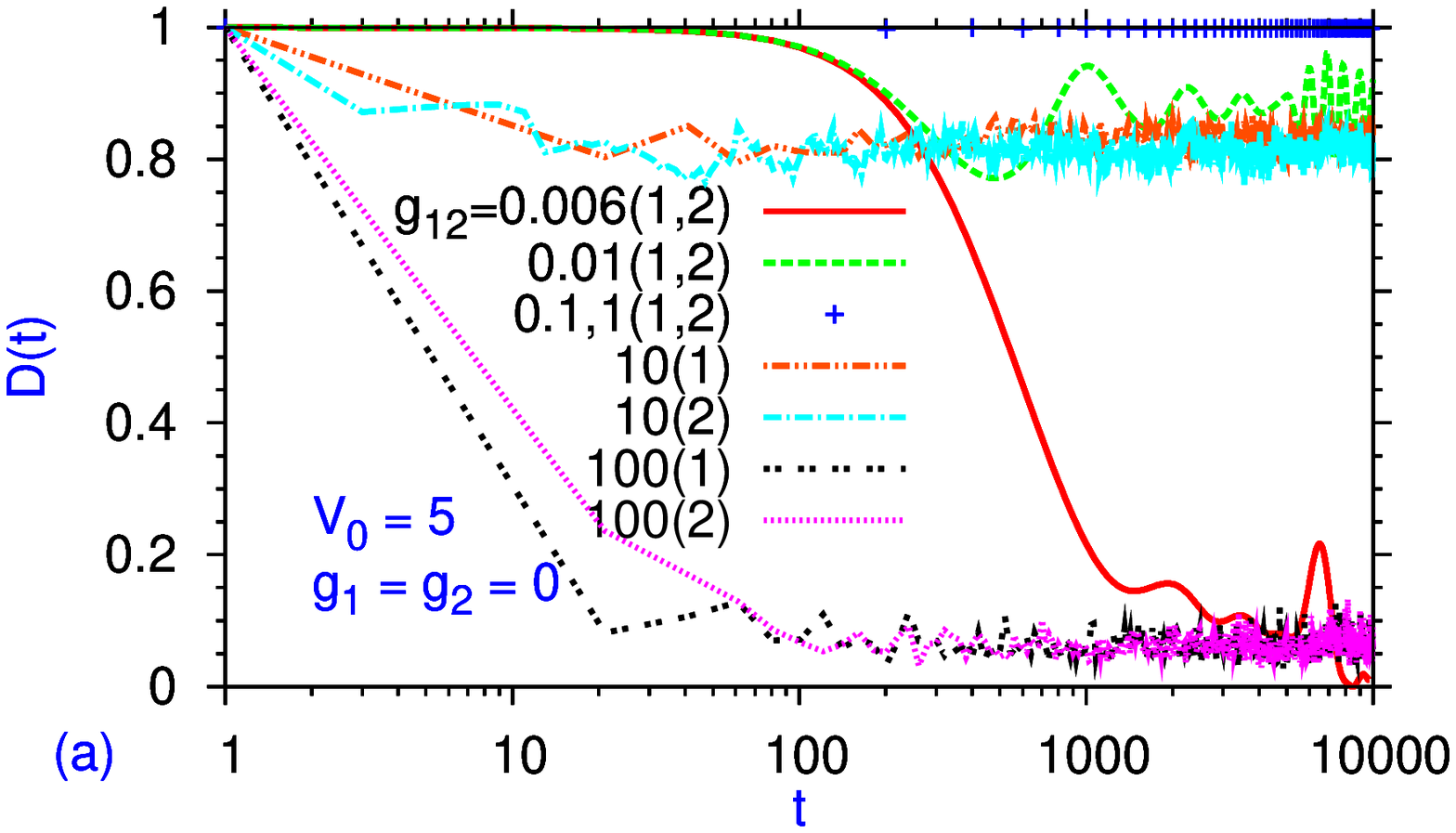}
\includegraphics[width=.49\linewidth]{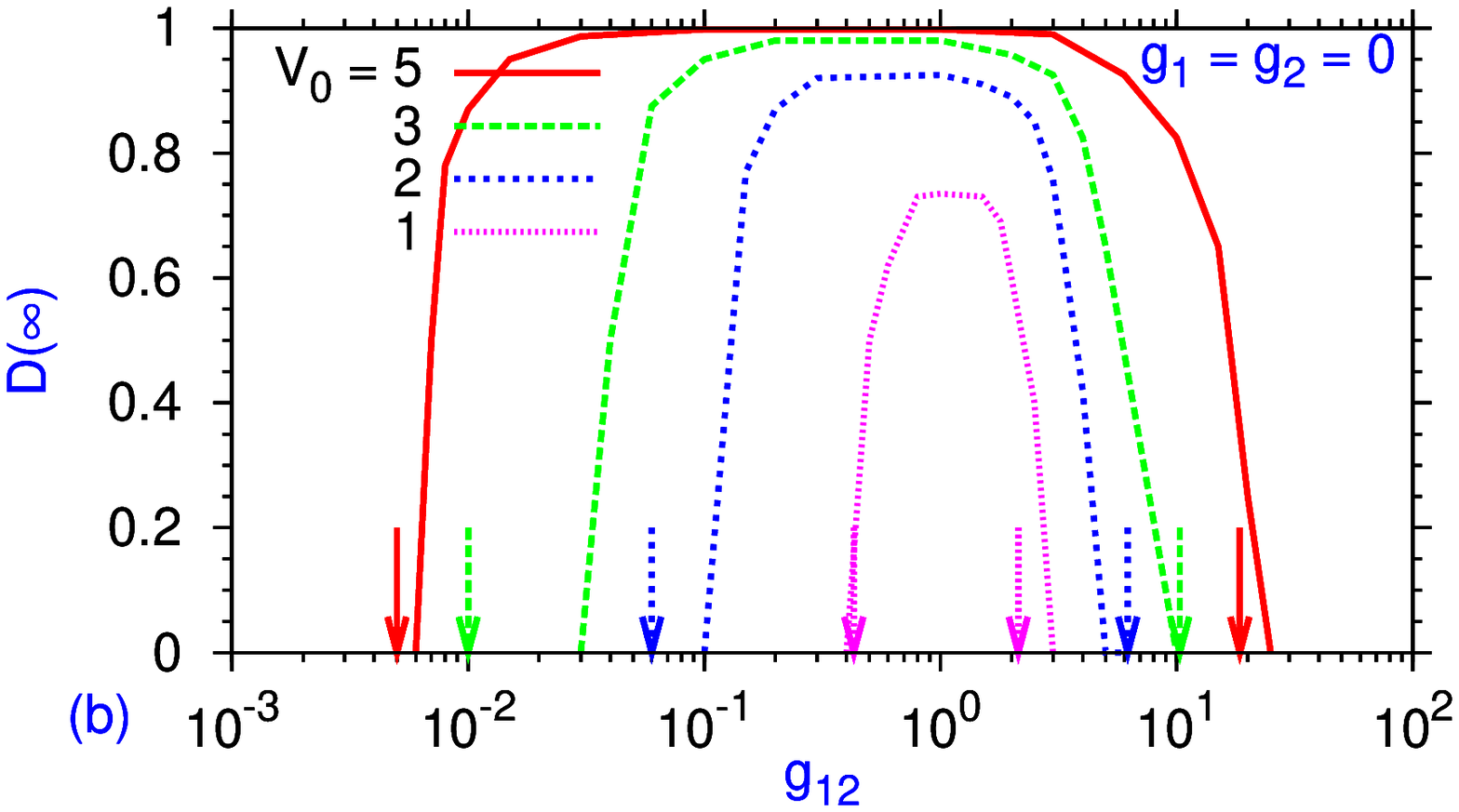}
\caption{
(a) Trapping measure $D(t)$ of a binary BEC
versus time $t$ for $V_0=5$ 
and different $g_{12}$. Curves are labeled by $g_{12}$ and component number.
(b) Trapping  measure at large time $D(\infty)$ versus $g_{12}$  for 
different $V_0$. 
The arrows represent  nonlinearities $g_{c1}$ and $g_{c2}$  from 
figure \ref{fig1} (b).
}
\label{fig9}
\end{figure}

{
Finally, we consider a binary BEC with small 
intraspecies nonlinearities  and zero interspecies 
nonlinearity ($g_{12}=0$), that does not allow self trapping, as found 
in figures \ref{fig7} (a) and (b). As an illustration we consider the 
zero values for the intraspecies nonlinearities: $g_1=g_2=0$.}
In the presence of appropriate 
repulsive interspecies nonlinearity ($g_{12}>0$), one can have 
self trapping  as illustrated in figures \ref{fig9} (a) and 
(b). In figure \ref{fig9} (a) we plot the trapping measure $D(t)$ versus 
$t$  for different repulsive (positive) interspecies 
nonlinearity $g_{12}$ and $V_0=5, g_1=g_2=0$.  In this case $g_{12}$ 
plays the role of effective interaction. There is a window of $g_{12}$ 
values
 $g_{c1}<g_{12} <g_{c2}$ with $g_{c1}\approx 0.01$ and $g_{c2}\approx 10$, 
  where permanent self trapping can be 
achieved.  In figure \ref{fig9} (b) we plot 
large-time trapping measure $D(\infty)$ versus $g_{12}$ for different 
$V_0$ and $g_1=g_2=0$. Qualitatively, this plot is quite similar to 
those in figures \ref{fig7} (b) and \ref{fig8} (b) showing the universal 
nature of these plots.

\section{Summary and Discussion}
\label{IV}

We demonstrated that self trapping of a BEC or a binary BEC without 
interspecies interaction in OL and DW  occurs for a window of 
repulsive intraspecies
nonlinearity $g$  ($g_{c1}<g<g_{c2}$), where $g_{c1}$ and $g_{c2}$ 
depends on the trap parameters. 
For a binary BEC with the intraspecies nonlinearities 
outside this window, a self trapping can be induced by a non-zero 
interspecies nonlinearity $g_{12}$ such that the effective nonlinearities
fall in this window. For intraspecies nonlinearities 
$g_i$ below $g_{c1}$ ($g_i<g_{c1}$), this is achieved by an
appropriate  repulsive (positive) $g_{12}$. For intraspecies nonlinearities 
$g_i$ above $g_{c2}$ ($g_i>g_{c2}$), one can have self trapping by 
introducing an appropriate  attractive (negative)
$g_{12}$. In case of self trapping in an OL,
the permanently self trapped state represents breathing oscillation of 
a stable stationary gap soliton. On the other hand, the self trapping in a
DW is purely a dynamical phenomenon without any underlying stationary state. 
However, the self trapping of a BEC and a binary BEC in both DW and OL 
could be permanent. 

In previous studies of self trapping of a BEC, the existence of a lower 
limit $g_{c1}$ of nonlinearity was
noted \cite{GatiJPB2007,trappingPRL,ol1,ol2}. 
However, the disappearance of 
self trapping above an upper limit   was not realized. 
(A similar upper limit appeared in a study of a Fermi 
superfluid in a DW \cite{DW1}.)
In 
a previous study of self trapping in an OL, in contradiction 
with the present investigation, 
it was concluded that 
self trapping was always transient and should disappear at large $t$ 
\cite{ol1}.

{{
The principal findings of this critical study of self trapping
 are the following. (a) The 
self-trapped states in an OL potential are essentially the stationary 
gap solitons. The self trapping in DW potential is entirely dynamical in 
nature and there are no stationary states in this case, such as the gap 
solitons of the OL potential. (There is no periodic potential and no 
band and band gap, specially 
for the case of a DW with a shallow barrier, which we shall 
study here.) It was generally believed that self 
trapping in both OL and DW potentials are dynamical in nature. (b) Self 
trapping is stopped beyond an upper limit of interaction in both cases. 
(c) In case of coupled systems with inter-species interaction self 
trapping is possible for domain of intra-species atomic interaction 
where no self trapping is allowed in uncoupled systems.}}

In the experiment of self trapping in an OL and in related theoretical 
studies a localized state over tens of OL sites \cite{ol1,ol2} was 
considered in contrast to that in predominantly a single OL 
site. Such states extended over multiple OL sites could 
possibly be a combination of multiple compact gap solitons. 
In another study, such states have been suggested to be a 
new type of spatially extended state in the gap \cite{gapself}. The 
compact self trapped state on OL considered in this paper are different 
from the spatially extended states considered in other studies. 
Nevertheless, with the present experimental control over a BEC, it 
would be possible to study self trapping of the compact states on OL as 
considered in this paper.

\ack

FAPESP and CNPq (Brazil) provided partial support.

\vskip 0.5 cm
References

\end{document}